\documentclass[twocolumn,floatfix,superscriptaddress,showpacs]{revtex4}
\usepackage{graphicx}
\usepackage{amsmath}
\usepackage{bm}
\begin{document}

\title{Optimal spin-entangled electron-hole pair pump}
\author{C. W. J. Beenakker}
\affiliation{Instituut-Lorentz, Universiteit Leiden, P.O. Box 9506, 2300 RA
Leiden, The Netherlands}
\author{M. Titov}
\affiliation{Max-Planck-Institut f\"{u}r Physik komplexer Systeme,
N\"{o}thnitzer Str.\ 38, 01187 Dresden, Germany}
\author{B. Trauzettel}
\affiliation{Instituut-Lorentz, Universiteit Leiden, P.O. Box 9506, 2300 RA
Leiden, The Netherlands}
\date{February 2005}
\begin{abstract}
A nonperturbative theory is presented for the creation by an oscillating
potential of spin-entangled electron-hole pairs in the Fermi sea. In the weak
potential limit, considered earlier by Samuelsson and B\"{u}ttiker, the
entanglement production is much less than one bit per cycle. We demonstrate
that a strong potential oscillation can produce an average of one Bell pair per
two cycles, making it an efficient source of entangled flying qubits.
\end{abstract}
\maketitle

The quantum electron pump is a device that transfers electrons phase coherently
between two reservoirs at the same voltage, by means of a slowly oscillating
voltage on a gate electrode \cite{Swi99}. Special pump cycles exist that
transfer the charge in a quantized fashion, one $e$ per cycle
\cite{Shu00,Lev01,Avr01,Mak01}. Building on earlier proposals to stochastically
produce entangled electron-hole pairs in a Fermi sea out of equilibrium
\cite{Bee03,Sam04a}, Samuelsson and B\"{u}ttiker have proposed \cite{Sam04}
that a quantum pump could be used to create entangled Bell pairs in a
controlled manner, clocked by the gate voltage. Such a device could be a
building block of quantum computing designs using ballistic flying qubits in
nanowires or in quantum Hall edge channels \cite{Ber00,Ion01,Bar04}.

To find out how close one get to this ideal, one needs to go beyond the
perturbation theory of Ref.\ \cite{Sam04} --- in which the number of Bell pairs
per cycle is $\ll 1$. A nonperturbative theory of the quantum entanglement pump
is presented here. We show that the entanglement production is closely related
to the charge noise, to the extent that a noiseless pump produces no
entanglement. By maximizing the charge noise with spin-independent scattering
we calculate that a pump can produce, on average, 1 Bell pair every 2 cycles. A
deterministic spin-entangler \cite{Bee04}, being the analogue of a quantized
charge pump, would have an entanglement production of $1$ Bell pair per cycle,
so the optimal entanglement pump has one half the efficiency of a deterministic
entangler.

\begin{figure}
\includegraphics[width=8cm]{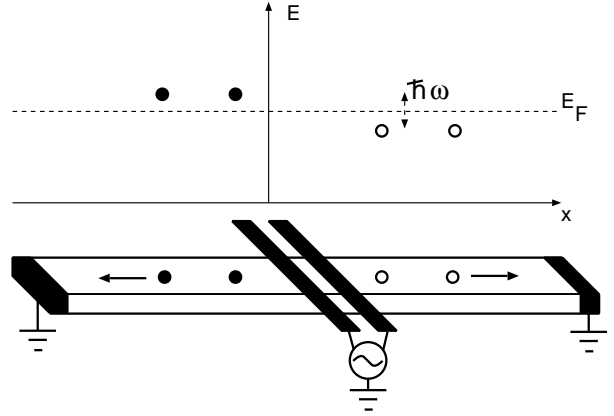}
\caption{
Production of entangled electron-hole pairs in a narrow ballistic conductor by
a quantum electron pump. The left and right ends of the conductor are at the
same potential, while the potential on the gate electrodes at the center is
periodically modulated. Such a device produces spatially separated
electron-hole pairs (black and white circles), differing in energy by a
multiple of the pump frequency $\omega$. For spin-independent scattering the
electron ($e$) and hole ($h$) produced during a given cycle have the same spin
$\uparrow,\downarrow$, so that their wave function is that of a Bell pair,
$\propto|\!\uparrow_{e}\uparrow_{h}\rangle+
|\!\downarrow_{e}\downarrow_{h}\rangle$. The optimal quantum entanglement 
pump produces, on average, $1$ Bell pair every $2$ cycles.
}
\label{epump}
\end{figure}

We consider a two-channel phase coherent conductor, see Fig.\ \ref{epump},
connecting a left and a right electron reservoir in thermal equilibrium (same
temperature $T$ and Fermi energy $E_{F}$ in each reservoir). The two channels
may refer to an orbital or to a spin degree of freedom. (To be definite, we
will usually speak of a spin degree of freedom.) A periodically varying
time-dependent electrical potential $V(\mathbf{r},t)$ (with period
$2\pi/\omega$) excites electron-hole pairs in the Fermi sea of the conductor.
The quantum mechanical state of an electron-hole pair, at energies $E,E'$
differing by a multiple of $\hbar\omega$, may be entangled in the channel
indices. The entanglement can be a resource for quantum computing if the
electron and the hole excitation are scattered to opposite ends of the
conductor, so that they become two separate qubits. We wish to relate this
entanglement production to the scattering matrix of the conductor.

The four characteristic energy scales of this problem are the thermal energy
$k_{B}T$, the pump energy $\hbar\omega$, the Thouless energy $\hbar/\tau_{D}$
(set by the inverse of the mean dwell time $\tau_{D}$ of an electron in the
conductor), and finally the Fermi energy $E_{F}$. In nanostructures at low
temperatures, the characteristic relative magnitude of these energy scales is
$k_{B}T\ll\hbar\omega\ll\hbar/\tau_{D} \ll E_{F}$. This is the adiabatic, low
temperature regime in which we will work.

We seek a relation between the entanglement production and the scattering
matrix $S$ of the pump, which is the unitary operator relating incoming to
outgoing states:
\begin{equation}
b_{n}(E)=\sum_{m}\int\frac{dE'}{2\pi}S_{nm}(E,E')a_{m}(E').\label{barelation}
\end{equation}
Here $a_{n}(E)$ is the fermion annihilation operator for an incoming channel
$n$ at energy $E$, and $b_{n}(E)$ is the annihilation operator for an outgoing
channel. There are four channels in total ($n=1,2,3,4$), two in the left lead
and two in the right lead. The Wigner transform of the scattering matrix
\cite{Vav01}, defined by
\begin{equation}
S_{W}(E,t)=\int\frac{dE'}{2\pi}\,S(E+E'/2,E-E'/2)e^{-iE't/\hbar},\label{SWdef}
\end{equation}
depends on $E$ on the scale $\hbar/\tau_{D}$. In the adiabatic regime
$\omega\tau_{D}\ll 1$ one may therefore neglect the $E$-dependence on the scale
of the pump energy, approximating $S_{W}(E,t)\approx S_{W}(E_{F},t)\equiv {\cal
S}(t)$. The $4\times 4$ unitary matrix ${\cal S}(t)$ can be obtained by solving
the static scattering problem for the frozen potential $V({\bf r},t)$.

Within a single period $2\pi/\omega$ the excitation energies can only be
resolved on the scale of $\hbar\omega$, so we discretize $E_{p}=p\hbar\omega$
with integer $p$. A pair of energies $E_{p}$ and $E_{q}$ is coupled by the
Floquet matrix ${\cal F}(p-q)$, which is the Fourier transform of the Wigner
transformed scattering matrix \cite{Mos02},
\begin{equation}
{\cal F}(p-q)=\frac{\omega}{2\pi}\int_{0}^{2\pi/\omega}dt\,{\cal
S}(t)e^{i(p-q)\omega t}.\label{Spqdef}
\end{equation}
The unitarity relation for the Floquet matrix reads
\begin{equation}
\sum_{n',p'}{\cal F}_{nn'}^{\vphantom{\ast}}(p-p'){\cal
F}_{mn'}^{\ast}(q-p')=\delta_{nm}\delta_{pq}.\label{FloquetFunitary}
\end{equation}

We assume zero temperature, so the incoming state $|\Psi_{0}\rangle$ is the
unperturbed Fermi sea, consisting of all levels (left and right) doubly
occupied below $E_{F}$ and empty above $E_{F}$:
\begin{equation}
|\Psi_{0}\rangle=\prod_{p}f(E_{p})\prod_{n}a^{\dagger}_{n}(E_{p})|0\rangle.
\label{Psiindef}
\end{equation}
The state $|0\rangle$ is the vacuum and $f(E)=\theta(E_{F}-E)$ is the
zero-temperature Fermi function. The outgoing state $|\Psi\rangle$ is obtained
from $|\Psi_{0}\rangle$ by substituting Eq.\ (\ref{barelation}) and taking the
adiabatic approximation for the scattering matrix,
\begin{equation}
|\Psi\rangle=\prod_{p}f(E_{p})\prod_{n}\biggl(\sum_{p'}\sum_{n'}b^{\dagger}_{n'}
(E_{p'}){\cal F}_{n'n}(p'-p)\biggr)|0\rangle.\label{Psioutdef}
\end{equation}

We denote by $w_{pq}^{eh}$ the probability that the pump excites within one
cycle a single electron-hole pair, consisting of an electron at the left at an
energy $E_{p}$ above the Fermi level and a hole at the right at an energy
$E_{q}$ below the Fermi level. The entanglement entropy (or entanglement of
formation) of the spins is denoted by ${\cal E}_{pq}^{eh}$ (measured in bits
per cycle). Similarly, $w_{qp}^{he}$ and ${\cal E}_{qp}^{he}$ refer to a hole
at the left and an electron at the right. The average production, per cycle, of
spin-entangled electron-hole pairs is
\begin{equation}
{\cal E}=\sum_{E_{p}>E_{F}}\sum_{E_{q}<E_{F}}\bigl(w_{pq}^{eh}{\cal
E}_{pq}^{eh}+w_{qp}^{he}{\cal E}_{qp}^{he}\bigr).\label{Etotal}
\end{equation}
A maximally entangled Bell pair has ${\cal E}_{pq}^{eh}=1$, so it contributes
$w_{pq}^{eh}$ bits to the entanglement production.

The weight factor $w_{pq}^{eh}$ and entanglement entropy ${\cal E}_{pq}^{eh}$
of an electron-hole pair can both be calculated by projecting $|\Psi\rangle$
onto a state ${\cal P}_{pq}^{eh}|\Psi\rangle$ with all levels (left and right)
empty above $E_{F}$ and doubly occupied below $E_{F}$ --- except for a singly
occupied level $E_{p}>E_{F}$ at the left and $E_{q}<E_{F}$ at the right. The
(unnormalized) projected electron-hole state has the form
\begin{equation}
{\cal
P}_{pq}^{eh}|\Psi\rangle=\sum_{\sigma,\sigma'=\uparrow,\downarrow}
\alpha_{\sigma\sigma'}b^{\dagger}_{L\sigma}(E_{p})b^{\dagger}_{R\sigma'}
(E_{q})|0\rangle.\label{Psitwoqubit}
\end{equation}
The four channels have been labeled
$L\!\!\uparrow,L\!\!\downarrow,R\!\!\uparrow,R\!\!\downarrow$, where $L,R$
refers to the left and right lead and the arrows $\uparrow,\downarrow$ indicate
the spin. The $2\times 2$ matrix $\alpha$ determines the weight factor as well
as the entanglement entropy,
\begin{eqnarray}
{\cal E}_{pq}^{eh}=-x\log_{2} x-(1-x)\log_{2}(1-x),\nonumber\\
\qquad x={\tfrac{1}{2}}+{\tfrac{1}{2}}\sqrt{1-{\cal C}^{2}},\label{calxdef}\\
w_{pq}^{eh}={\rm Tr}\,\alpha\alpha^{\dagger},\;\;{\cal C}=\frac{2({\rm
Det}\,\alpha\alpha^{\dagger})^{1/2}}{{\rm
Tr}\,\alpha\alpha^{\dagger}}.\label{wCdef}
\end{eqnarray}
The number ${\cal C}\in[0,1]$ is the concurrence \cite{Woo98} of the
electron-hole pair.

In order to calculate the matrix $\alpha$ it is more convenient to perform the
algebraic manipulations on the pair correlator $K$ in the outgoing state
$|\Psi\rangle$, rather than on the state itself. The pair correlator fully
characterizes the outgoing state (\ref{Psioutdef}) because it is Gaussian,
meaning that higher order correlators in normal order (all $b^{\dagger}$'s to
the left of the $b$'s) are constructed from the pair correlator according to
the rule of Gaussian averages. The correlator is given in terms of the Floquet
matrix by
\begin{eqnarray}
&&K_{nm}(p,q)=\langle b_{m}^{\dagger}(E_{q})b_{n}^{\vphantom{\dagger
}}(E_{p})\rangle\nonumber\\
&&\qquad=\sum_{n',p'}{\mathcal F}_{nn'}^{\vphantom{\ast}}
(p-p')f(E_{p'}){\mathcal F}_{mn'}^{\ast}(q-p').\label{KFrelation}
\end{eqnarray}
The matrix $K$ is Hermitian and idempotent in the joint set of energy and
channel indices: $K=K^{\dagger}=K^{2}$. This signifies that the state it
represents is a pure (rather than a mixed) state \cite{note2}.

Projection of $|\Psi\rangle$ onto a set of filled or empty levels preserves the
Gaussian property. The correlator $\bar{K}$ of the projected state ${\cal
P}_{pq}|\Psi\rangle$ is derived from $K$ by the procedure known in matrix
algebra as Gaussian elimination \cite{Eis02}. By interchanging rows and columns
of the matrix $K$ we move the indices
$pL\!\!\uparrow,pL\!\!\downarrow,qR\!\!\uparrow,qR\!\!\downarrow$ to the upper
left hand corner, to obtain the block form
\begin{equation}
K=\begin{pmatrix}
K_{\rm dir}&X\\
X^{\dagger}&Y
\end{pmatrix},\;
K_{\rm dir}=
\begin{pmatrix}
K_{LL}(p,p)&K_{LR}(p,q)\\
K_{RL}(q,p)&K_{RR}(q,q)
\end{pmatrix}.\label{Kdecomposed}
\end{equation}
The $4\times 4$ block $K_{\rm dir}$ contains the direct coupling of the spin
degenerate levels $p$ at the left and $q$ at the right. The correlator
$\bar{K}$ contains in addition the indirect coupling via the filled or empty
states in the block $Y$,
\begin{subequations}
\label{barKbarw}
\begin{eqnarray}
&&\bar{K}=K_{\rm dir}+X(1-Y-\Lambda)^{-1}X^{\dagger},\label{KSchur}\\
&&\bar{w}\equiv\langle\Psi|{\cal P}_{pq}|\Psi\rangle=|{\rm
Det}\,(1-Y-\Lambda)|.\label{barwdef}
\end{eqnarray}
\end{subequations}
The diagonal matrix $\Lambda$ has a $1$ on the diagonal if the state is filled
(below $E_{F}$) and a $0$ if it is empty (above $E_{F}$). A derivation of Eq.\
(\ref{barKbarw}) is given in App.\ \ref{appA}.

One readily verifies that $\bar{K}^{2}=\bar{K}$, so the projection preserves
the purity of the state, as it should. Since ${\cal P}_{pq}|\Psi\rangle$
contains a total of two electrons in four states, the correlator $\bar{K}$ has
two eigenvalues equal to 1 and two eigenvalues equal to 0. We write
$\bar{K}=U{\rm diag}\,(0,0,1,1)U^{\dagger}$, with $U$ the unitary matrix of
eigenvectors. The projected state corresponding to the correlator $\bar{K}$ has
the form
\begin{eqnarray}
&&{\cal P}_{pq}|\Psi\rangle=\bar{w}^{1/2}({\bf b}^{\dagger}U)_{R\uparrow}({\bf
b}^{\dagger}U)_{R\downarrow}|0\rangle,\label{barPsi}\\
&&{\bf
b}=\bigl(b_{L\uparrow}(E_{p}),b_{L\downarrow}(E_{p}),b_{R\uparrow}(E_{q}),
b_{R\downarrow}(E_{q})\bigr).\label{bbolddef}
\end{eqnarray}
The matrix $U$ plays the role of an effective scattering matrix for the two
electrons in the two states left and right, including in addition to the direct
scattering (described by the original scattering matrix $S$) also the indirect
transitions via the other states.

To obtain the required projection ${\cal P}_{pq}^{eh}|\Psi\rangle$ we still
need to project ${\cal P}_{pq}|\Psi\rangle$ onto a state with a single electron
left and a single hole right,
excluding the double occupation. (We could not do the projection in a single
step because the final state (\ref{Psitwoqubit}) is not Gaussian, so it can not
be represented by a pair correlator.) By comparing Eqs.\ (\ref{Psitwoqubit})
and (\ref{barPsi}) we can relate the coefficient matrices $U$ and $\alpha$
before and after projection,
\begin{equation}
\alpha=i\bar{w}^{1/2}U_{LR}\sigma_{y}U_{RR}^{T},\;\;
U=\begin{pmatrix}
U_{LL}&U_{LR}\\
U_{RL}&U_{RR}
\end{pmatrix}.\label{alpharesult}
\end{equation}
(The matrix $\sigma_{y}$ is a Pauli matrix.) Substitution into Eqs.\
(\ref{calxdef}) and (\ref{wCdef}) then gives the contribution from this
electron-hole pair to the entanglement production.

A major simplification occurs in the case of spin-independent scattering. Then
$U_{LR}$ and $U_{RR}$ are proportional to the $2\times 2$ unit matrix
$\openone$, so $\alpha\propto\sigma_{y}$ and the electron-hole pair is
maximally entangled (${\cal E}_{pq}^{eh}=1$). In view of Eq.\ (\ref{Etotal})
the average entanglement production per cycle,
\begin{equation}
{\cal
E}=\sum_{E_{p}>E_{F}}\sum_{E_{q}<E_{F}}\bigl(w_{pq}^{eh}+w_{qp}^{he}\bigr),
\label{Etotalnospin}
\end{equation}
is the probability that the pump produces a single spatially separated
electron-hole pair in a given cycle.

The probability (\ref{Etotalnospin}) can be rewritten as ${\cal
E}=P_{0}^{\uparrow}P_{1}^{\downarrow}+P_{0}^{\downarrow}P_{1}^{\uparrow}$,
where $P_{\nu}^{\sigma}$ is the probability that $\nu$ spatially separated
electron-hole pairs of spin $\sigma$ are produced in a given cycle. From $0\leq
P_{1}^{\uparrow}=P_{1}^{\downarrow}\leq
1-P_{0}^{\uparrow}=1-P_{0}^{\downarrow}\leq 1$ we deduce that
\begin{equation}
{\cal E}\leq 2P_{0}^{\sigma}(1-P_{0}^{\sigma})\leq \tfrac{1}{2}.\label{Emax}
\end{equation}
This maximal entanglement ${\cal E}_{\rm max}=\frac{1}{2}$ of one half bit per
cycle is reached for
$P_{0}^{\uparrow}=P_{0}^{\downarrow}=P_{1}^{\uparrow}=P_{1}^{\downarrow}=
\frac{1}{2}$. Eq.\ (\ref{Emax}) is derived for spin-independent scattering. 
It seems unlikely that spin-dependent scattering (which reduces the entanglement 
per electron-hole pair) could violate the bound ${\cal E}\leq\frac{1}{2}$, 
but we have not been able to exclude this possibility on mathematical grounds.

To demonstrate that the optimal value ${\cal E}_{\rm max}=\frac{1}{2}$ can be
reached, we consider the pump cycle
\begin{equation}
{\cal S}(\tau)=\begin{pmatrix}
e^{i\omega\tau}r&t\\t'&e^{-i\omega\tau}r'
\end{pmatrix},
\label{Stsimple}
\end{equation}
which has been used as a model for a quantized charge pump \cite{Shu00,And00}.
(A more general class of pump cycles \cite{Iva97} with the same entanglement
production is analyzed in App.\ \ref{appB}.) Choosing the Fermi level such that
$E_{0}<E_{F}<E_{1}$, Eq.\ (\ref{KFrelation}) evaluates to
\begin{eqnarray}
K(p,q)&=&\begin{pmatrix}\openone&0\\0&\openone\end{pmatrix}
\delta_{pq}f(E_{p})\nonumber\\
&&\mbox{}+
\begin{pmatrix}-rr^{\dagger}\delta_{p0}\delta_{q0}&tr'^{\dagger}
\delta_{p0}\delta_{q1}\\r't^{\dagger}\delta_{p1}\delta_{q0}&r'r'^{\dagger}
\delta_{p1}\delta_{q1}\end{pmatrix}.\label{Ksimple}
\end{eqnarray}
The only pair of coupled levels is $E_{0}$ at the left and $E_{1}$ at the
right, so the entanglement production consists of a single term ${\cal
E}=w_{01}^{he}{\cal E}_{01}^{he}$. The matrix $X$ in the decomposition
(\ref{Kdecomposed}) vanishes, and $Y=\Lambda$, hence Eq.\ (\ref{barKbarw})
simplifies to
\begin{equation}
\bar{K}=K_{\rm
dir}=\begin{pmatrix}\openone-rr^{\dagger}&tr'^{\dagger}\\
r't^{\dagger}&r'r'^{\dagger}\end{pmatrix},\;\;\bar{w}=1.\label{Kbarsimple}
\end{equation}
Eq.\ (\ref{alpharesult}) gives $\alpha=it\sigma_{y}r'^{T}$, which finally leads
to the entanglement production
\begin{equation}
{\cal E}=H(x_{1},x_{2}),\;\;
x_{1}=T_{1}(1-T_{2}),\;\;x_{2}=T_{2}(1-T_{1}),\label{Esimple}
\end{equation}
in terms of the function
\begin{equation}
H(x,y)=(x+y)\log_{2}(x+y)-x\log_{2}x-y\log_{2}y\label{Hfundef}
\end{equation}
of the two transmission eigenvalues $T_{1},T_{2}$ (eigenvalues of
$tt^{\dagger}$, equal to the eigenvalues of $t't'^{\dagger}$ because of
unitarity of ${\cal S}$).

The optimal entanglement production ${\cal E}_{\rm max}=\frac{1}{2}$ is reached
in Eq.\ (\ref{Esimple}) for $T_{1}=T_{2}=\frac{1}{2}$ (corresponding to
spin-independent scattering, as expected). This is also the choice of
parameters at which the charge noise $\propto T_{1}(1-T_{1})+T_{2}(1-T_{2})$ is
maximized \cite{And00}. Although entanglement entropy and charge noise are
different physical quantities, with a different dependence on the transmission
eigenvalues, quite generally one can state that there can be no entanglement
production without charge noise. Indeed, a deterministic spin-independent
charge pump has $P_{0}^{\sigma}=0$ hence ${\cal E}=0$, in view of Eq.\
(\ref{Emax}).

A one-to-one relationship between entanglement production and charge noise is
possible in the weak pumping limit of Ref.\ \cite{Sam04}. To demonstrate this,
we quantify the pumping strength by a dimensionless parameter $\epsilon\ll 1$,
and calculate both quantities to leading order in $\epsilon$. The Floquet
matrix to first order has the general form
\begin{equation}
{\cal F}(p-q)=\left\{\begin{array}{ll}
{\cal F}_{0}&{\rm if}\;\;p=q,\\
i\epsilon Q{\cal F}_{0}&{\rm if}\;p=q+1,\\
i\epsilon Q^{\dagger}{\cal F}_{0}&{\rm if}\;p=q-1.
\end{array}\right.\label{Fweak}
\end{equation}
Unitarity of ${\cal F}_{0}$ ensures unitarity of ${\cal F}$ up to terms of
order $\epsilon^{2}$. The corresponding correlator (\ref{KFrelation}) is
\begin{eqnarray}
K(p,q)&=&f(E_{p})\delta_{pq}+i\epsilon
[f(E_{q})-f(E_{p})]\delta_{p,q+1}Q\nonumber\\
&&\mbox{}-i\epsilon [f(E_{p})-f(E_{q})]\delta_{q,p+1}Q^{\dagger}.\label{Kweak}
\end{eqnarray}

Following the same steps as before, we arrive at the entanglement production
\begin{equation}
{\cal
E}=\epsilon^{2}H\bigl(y_{1},y_{2}\bigr)+\epsilon^{2}
H\bigl(y'_{1},y'_{2}\bigr),\label{Eweak}
\end{equation}
in terms of the function $H$ defined in Eq.\ (\ref{Hfundef}). The numbers
$y_{n}$ and $y'_{n}$ are the eigenvalues of the matrices $\tau\tau^{\dagger}$
and $\tau'\tau'^{\dagger}$, respectively, constructed from sub-blocks of the
matrix
\begin{equation}
Q=\begin{pmatrix}
\rho&\tau\\
\tau'&\rho
\end{pmatrix}.\label{Qblock}
\end{equation}
In the case of spin-independent scattering $y_{1}=y_{2}\equiv y$,
$y'_{1}=y'_{2}\equiv y'$, and Eq.\ (\ref{Eweak}) simplifies to
\begin{equation}
{\cal E}=2\epsilon^{2}(y+y').\label{Eweaknospin}
\end{equation}

To compare this result with the charge noise ${\cal P}$, we use the formula
\cite{Mos02,And00,Pol02,note1}
\begin{eqnarray}
&&{\cal P}=\frac{1}{4}\sum_{p=1}^{\infty}p\,{\rm
Tr}\,G(p)G(-p),\label{Pformula}\\
&&G(p)=\sum_{q=-\infty}^{\infty}{\cal F}^{\dagger}(q-p)\begin{pmatrix}
\openone&0\\
0&-\openone
\end{pmatrix}
{\cal F}(q).\label{Gpdef}
\end{eqnarray}
In the weak pumping limit this reduces to
\begin{equation}
{\cal P}=\epsilon^{2}\bigl(y_{1}+y_{2}+y'_{1}+y'_{2}\bigr),\label{Presult}
\end{equation}
which equals the entanglement production (\ref{Eweaknospin}) in the
spin-independent case.

The close relation between entanglement production and charge noise in the weak
pumping regime is consistent with the finding of Ref.\ \cite{Sam04} that
low-frequency noise measurements can be used to detect the entanglement in this
regime. To access the nonperturbative regime investigated in this paper
requires time-resolved detection, on the time scale of $1/\omega$. The
requirement that the thermal energy $k_{B}T$ remains less than $\hbar\omega$
poses a practical lower limit to the frequency. What motivates further efforts
on the side of the detection is the relative simplicity on the side of the
production: The quantum entanglement pump requires no advanced lithography or
control over electron-electron interactions to produce as much as one Bell pair
per two cycles.

This work was supported by the Dutch Science Foundation NWO/FOM and by the U.S.
Army Research Office (Grant No.\ DAAD 19--02--0086).

\appendix
\section{Gaussian elimination of fermion degrees of freedom}
\label{appA}

The Gaussian elimination of boson degrees of freedom was done in Ref.\
\cite{Eis02}. Here we perform the analogous calculation for fermions, leading
to Eq.\ (\ref{barKbarw}). We represent the density matrix $\rho$ by the
normally ordered characteristic function
\begin{equation}
\chi(\bm{\xi})={\rm Tr}\,\rho\,
e^{\bm{\xi}\cdot\bm{b}^{\dagger}}e^{\bm{\xi}^{\ast}\cdot\bm{b}}.\label{chidef}
\end{equation}
The entire set of fermion annihilation operators is contained in the vector
$\bm{b}$. The characteristic function depends on the vector of Grassmann
variables $\bm{\xi}$. For a Gaussian state, with pair correlator $\bm{K}$, one
has
\begin{equation}
\chi(\bm{\xi})=\exp(\bm{\xi}^{\ast}\cdot{\bm
K}\cdot\bm{\xi}).\label{chiGaussian}
\end{equation}

We first eliminate a single degree freedom, labeled by a subscript $0$, by
projecting onto a filled state. The projection operator is
$b_{0}^{\dagger}b_{0}^{\vphantom{\dagger}}$, so the characteristic function
after projection is
\begin{eqnarray}
\chi_{0,{\rm filled}}(\bm{\xi})&=&{\rm
Tr}\,b_{0}^{\dagger}b_{0}^{\vphantom{\dagger}}\rho\,
e^{\bm{\xi}\cdot\bm{b}^{\dagger}}e^{\bm{\xi}^{\ast}\cdot\bm{b}}\nonumber\\
&=&\lim_{\xi_{0},\xi_{0}^{\ast}\rightarrow
0}\frac{d}{d\xi_{0}}\frac{d}{d\xi_{0}^{\ast}}\chi(\bm{\xi})\nonumber\\
&=&\int d\xi_{0}\int d\xi_{0}^{\ast}\,\chi(\bm{\xi}).\label{chifilled}
\end{eqnarray}
The final equality holds because differentiation and integration is the same
for Grassmann variables. Similarly, we can project onto an empty state, with
projection operator $1-b_{0}^{\dagger}b_{0}^{\vphantom{\dagger}}$ and
characteristic function
\begin{eqnarray}
\chi_{0,{\rm empty}}(\bm{\xi})&=&{\rm
Tr}\,(1-b_{0}^{\dagger}b_{0}^{\vphantom{\dagger}})\rho\,
e^{\bm{\xi}\cdot\bm{b}^{\dagger}}e^{\bm{\xi}^{\ast}\cdot\bm{b}}\nonumber\\
&=&-\lim_{\xi_{0},\xi_{0}^{\ast}\rightarrow
0}\frac{d}{d\xi_{0}}\frac{d}{d\xi_{0}^{\ast}}e^{-\xi_{0}^{\ast}
\xi_{0}^{\vphantom{\ast}}}\chi(\bm{\xi})\nonumber\\
&=&-\int d\xi_{0}\int
d\xi_{0}^{\ast}\,e^{-\xi_{0}^{\ast}\xi_{0}^{\vphantom{\ast}}}
\chi(\bm{\xi}).\label{chiempty}
\end{eqnarray}

We now divide the degrees of freedom into two sets, those that are to be
eliminated by projecting onto filled or empty states (labeled $P$), and those
that are to be retained (labeled $R$). The diagonal matrix $\bm{\Lambda}$, in
the space of projected states, has a $0$ on the diagonal if the state is empty
and a $1$ if the state is filled. The characteristic function after projection
is
\begin{eqnarray}
\chi(\bm{\xi}_{R})&=&{\rm Tr}\,\prod_{n\in
P}[1-\Lambda_{nn}-(-1)^{\Lambda_{nn}}b_{n}^{\dagger}b_{n}^{\vphantom{\dagger}}]
\rho\, e^{\bm{\xi}\cdot\bm{b}^{\dagger}}e^{\bm{\xi}^{\ast}\cdot\bm{b}}\nonumber\\
&=&\prod_{n\in P}(-1)^{1-\Lambda_{nn}}\int d\xi_{n}\int
d\xi_{n}^{\ast}\nonumber\\
&&\mbox{}\times e^{-\bm{\xi}_{P}^{\ast}\cdot(1-{\bm
\Lambda})\cdot\bm{\xi}_{P}^{\vphantom{\ast}}}\chi(\bm{\xi}).\label{chiR}
\end{eqnarray}

The Gaussian characteristic function (\ref{chiGaussian}) has the block
structure
\begin{equation}
\chi(\bm{\xi}_{R},\bm{\xi}_{P})=\exp\left[
\begin{pmatrix}
\bm{\xi}_{R}^{\ast}\\
\bm{\xi}_{P}^{\ast}
\end{pmatrix}
\cdot
\begin{pmatrix}
{\bm K}_{\rm dir}&{\bm X}\\
{\bm X}^{\dagger}&{\bm Y}
\end{pmatrix}
\cdot
\begin{pmatrix}
\bm{\xi}_{R}\\
\bm{\xi}_{P}
\end{pmatrix}
\right].\label{chidecomposed}
\end{equation}
Substitution into Eq.\ (\ref{chiR}) gives, upon integration over the Grassman
variables,
\begin{eqnarray}
&&\chi(\bm{\xi}_{R})=\prod_{n\in P}(-1)^{1-\Lambda_{nn}}\,{\rm Det}\,({\bm
Y}-1+\bm{\Lambda})\nonumber\\
&&\;\;\mbox{}\times\exp\biggl[\bm{\xi}_{R}^{\ast}\cdot\bigl(\bm{K}_{\rm
dir}-\bm{X}\cdot(\bm{Y}-1+\bm{\Lambda})^{-1}\cdot\bm{X}^{\dagger}\bigr)
\cdot\bm{\xi}_{R}\biggr].\nonumber\\
&&\label{chiRresult}
\end{eqnarray}
This is the characteristic function of a Gaussian state with weight and pair
correlator given by Eq.\ (\ref{barKbarw}). We note that this derivation holds
regardless of whether $\rho$ represents a pure or a mixed state (so regardless
of whether $K^{2}=K$ or not).

\begin{widetext}
\section{Class of optimal entanglement pump cycles}
\label{appB}

We calculate the entanglement production for pump cycles of the form
\begin{equation}
{\cal S}(\tau)=\begin{pmatrix}
e^{i\vartheta(\tau)}r&t\\t'&e^{-i\vartheta(\tau)}r'
\end{pmatrix},\;\;
e^{i\vartheta(\tau)}=\frac{e^{i\omega\tau}-z}{1-z^{\ast}e^{i\omega\tau}},
\label{Stz}
\end{equation}
introduced by Ivanov, Lee, and Levitov \cite{Iva97} in the context of charge
pumps. The complex parameter $z$ satisfies $|z|^2< 1$. We will show that these
pump cycles all have the same entanglement production as the special case $z=0$
considered in the main text [Eqs.\ (\ref{Stsimple})--(\ref{Hfundef})]. In
particular, they are optimal for any $z$ if $T_{1}=T_{2}=\frac{1}{2}$.

We label the energy levels such that $E_{p}<E_{F}$ if $p\leq 0$ and
$E_{p}>E_{F}$ if $p\geq 1$. The pair correlator is given by
\begin{equation}
K(p,q)=\begin{pmatrix}\openone&0\\0&\openone\end{pmatrix}
\delta_{pq}f_{p}+(1-|z|^2)
\begin{pmatrix}
-rr^{\dagger}f_{p}f_{q}z^{-q}(z^{\ast})^{-p} & tr'^{\dagger}
f_{p}(1-f_{q})(z^{\ast})^{q-p-1}\\
r't^{\dagger}(1-f_{p})f_{q}z^{p-q-1} &
r'r'^{\dagger}(1-f_{p})(1-f_{q})z^{p-1}(z^{\ast})^{q-1}
\end{pmatrix}.\label{Kz}
\end{equation}
We have abbreviated $f_{p}\equiv f(E_{p})$. This pump transfers electrons
exclusively from left to right, so only half of the terms in the entanglement
production (\ref{Etotal}) need to be included:
\begin{equation}
{\cal E}=\sum_{p=-\infty}^{0}\sum_{q=1}^{\infty}w_{pq}^{he}{\cal
E}_{pq}^{he}.\label{Etotalz}
\end{equation}
The index $p\leq 0$ refers to a hole excitation at the left and the index
$q\geq 1$ to an electron excitation at the right.

The Gaussian elimination (\ref{barKbarw}) is more complicated than for the case
$z=0$ considered in the main text, because the coupling between two given
levels $p$ and $q$ now contains also an indirect contribution (via an infinite
ladder of hole excitations at the left and electron excitations at the right).
Still, after some algebra the final result for the correlator $\bar{K}$ of the
projected state ${\cal P}_{pq}|\Psi\rangle$ turns out to be rather simple. We
introduce the polar decomposition
\begin{equation}
\begin{pmatrix}
r&t\\t'&r'
\end{pmatrix}
=
\begin{pmatrix}
u&0\\0&v
\end{pmatrix}
\begin{pmatrix}
\sqrt{1-T}&-\sqrt{T}\\
\sqrt{T}&\sqrt{1-T}
\end{pmatrix}
\begin{pmatrix}
u'&0\\0&v'
\end{pmatrix},\label{polar}
\end{equation}
where $u,v,u',v'$ are $2\times 2$ unitary matrices and $T={\rm
diag}\,(T_{1},T_{2})$ is a diagonal matrix of transmission eigenvalues.
Defining also $c_{pq}=(1-|z|^{2})z^{q-p-1}$, the result for $\bar{K}$ can be
written in the form
\begin{equation}
\bar{K}=\begin{pmatrix}
u&0\\0&v
\end{pmatrix}
\begin{pmatrix}
T+(1-T)|c_{pq}|^{2}&0\\
0&T+(1-T)|c_{pq}|^{2}
\end{pmatrix}^{-1}
\begin{pmatrix}
T&-c_{pq}^{\ast}\sqrt{T(1-T)}\\-c_{pq}\sqrt{T(1-T)}&(1-T)|c_{pq}|^{2}
\end{pmatrix}
\begin{pmatrix}
u^{\dagger}&0\\0&v^{\dagger}
\end{pmatrix}.\label{barKz}
\end{equation}
The weight factor $\bar{w}$ equals
\begin{equation}
\bar{w}=\prod_{n=1,2}\left[T_{n}+(1-T_{n})|c_{pq}|^{2}\right].\label{barwz}
\end{equation}

The unitary matrix $U$ that diagonalizes $\bar{K}=U{\rm
diag}\,(0,0,1,1)U^{\dagger}$ is given by
\begin{equation}
U=
\begin{pmatrix}
u&0\\0&v
\end{pmatrix}
\begin{pmatrix}
T+(1-T)|c_{pq}|^{2}&0\\
0&T+(1-T)|c_{pq}|^{2}
\end{pmatrix}^{-1/2}
\begin{pmatrix}
c_{pq}^{\ast}\sqrt{1-T}&\sqrt{T}\\
\sqrt{T}&-c_{pq}\sqrt{1-T}
\end{pmatrix},\label{Uz}
\end{equation}
hence Eq.\ (\ref{alpharesult}) gives
\begin{equation}
\alpha=-i\bar{w}^{1/2}c_{pq}u\left(\frac{T}{T+(1-T)|c_{pq}|^{2}}\right)^{1/2}
\sigma_{y}\left(\frac{1-T}{T+(1-T)|c_{pq}|^{2}}\right)^{1/2}v^{T}.\label{alphaz}
\end{equation}
\end{widetext}

Substitution of Eq.\ (\ref{alphaz}) into Eqs.\ (\ref{calxdef}) and
(\ref{wCdef}) gives the weighted entanglement production of the electron-hole
pair,
\begin{equation}
w_{pq}^{he}{\cal
E}_{pq}^{he}=|c_{pq}|^{2}H\bigl(T_{1}(1-T_{2}),T_{2}(1-T_{1})\bigr),\label{wEz}
\end{equation}
in terms of the function $H$ defined in Eq.\ (\ref{Hfundef}). Finally, we carry
out the sum over the level indices $p$ and $q$ in Eq.\ (\ref{Etotalz}) to
obtain the total entanglement production
\begin{equation}
{\cal E}=H\bigl(T_{1}(1-T_{2}),T_{2}(1-T_{1})\bigr),\label{Ezfinal}
\end{equation}
which is the same as Eq. (\ref{Esimple}). The $z$-dependence drops out because
$\sum_{p=-\infty}^{0}\sum_{q=1}^{\infty}|c_{pq}|^{2}=1$. The maximum ${\cal
E}=\frac{1}{2}$ is reached for $T_{1}=T_{2}=\frac{1}{2}$, independent of $z$.


\begin{thebibliography}{99}
\bibitem{note1} The charge noise ${\cal P}$ is defined by ${\rm Var}\,{\cal
Q}=(e^{2}\omega t_{\rm det}/2\pi){\cal P}$, in terms of the variance of the
charge ${\cal Q}$ transferred in the detection time $t_{\rm det}\gg 1/\omega$.
\end{thebibliography}
\end{document}